\documentstyle[12pt]{article}
\begin{document}
\newcommand{\ket}[1] {\mbox{$ \vert #1 \rangle $}}
\newcommand{\bra}[1] {\mbox{$ \langle #1 \vert $}}
\newcommand{\bkn}[1] {\mbox{$ < #1 > $}}
\newcommand{\bk}[1] {\mbox{$ \langle #1 \rangle $}}
\newcommand{\scal}[2]{\mbox{$ \langle  #1 \vert #2 \rangle   $}}
\newcommand{\expect}[3] {\mbox{$ \bra{#1} #2 \ket{#3} $}}
\newcommand{\ki}{\mbox{$ \ket{\psi_i} $}}
\newcommand{\bi}{\mbox{$ \bra{\psi_i} $}}
\newcommand{\p} \prime
\newcommand{\e} \epsilon
\newcommand{\la} \lambda
\newcommand{\om} \omega   \newcommand{\Om} \Omega
\newcommand{\cc}{\mbox{$\cal C $}}
\newcommand{\w} {\hbox{ weak }}
\newcommand{\al} \alpha
\newcommand{\m} {\mbox{$\mu_k$}}
\newcommand{\n} {\mbox{$\nu_k$}}
\newcommand{\bt} \beta
\newcommand{\La} \Lambda 

\newcommand{\be} {\begin{equation}}
\newcommand{\ee} {\end{equation}}
\newcommand{\ba} {\begin{eqnarray}}
\newcommand{\ea} {\end{eqnarray}}

\def\lrD{\mathrel{{\cal D}\kern-1.em\raise1.75ex\hbox{$\leftrightarrow$}}}
\def\lr #1{\mathrel{#1\kern-1.25em\raise1.75ex\hbox{$\leftrightarrow$}}}
\newcommand{\lrpartial}[1] {\mbox{$\stackrel{\leftrightarrow}
{\partial_{ #1}}$}}
%\mathrel
%{\partial\kern-.75em\raise1.75ex\hbox{$\leftrightarrow$}}}
%Serge 2-3-99

\centerline{\LARGE\bf{
Unitary and Non-Unitary Evolution 
 }}\vskip .3 truecm
\centerline{\LARGE\bf{in Quantum Cosmology}}
\vskip 1.5 truecm
\centerline{{\bf S. Massar}}
\centerline{Service de Physique Th\'eorique, 
Universit\'e Libre de Bruxelles,}
\centerline{
CP 225, Bvd du Triomphe, B1050 Brussels, 
Belgium.}
\vskip .5 truecm
\centerline{and}
\vskip .5 truecm
\centerline{{\bf R. Parentani}}
\centerline{Laboratoire de Math\'ematiques et Physique Th\'eorique,
CNRS UPRES A 6083,}
\centerline{Facult\' e des Sciences, 
Universit\'e de Tours, 37200 Tours, France. }
\vskip 1.5 truecm
\centerline{\bf{abstract}}

We analyze when and why unitarity violations might occur
in quantum cosmology restricted to minisuperspace. To this end we 
discuss in detail
backscattering transitions between expanding and contracting solutions of the 
Wheeler-DeWitt equation.  We first show that upon neglecting only
backscattering, one obtains an intermediate regime in which matter evolves 
unitarily
but which does not correspond to any Schr\"odinger equation in a given 
geometry
since gravitational backreaction effects are taken into account at the quantum 
level. 
We then show that backscattering amplitudes are 
exponentially smaller than matter transition amplitudes. 
Both results follow from an adiabatic treatment valid for macroscopic 
universes.
To understand how backscattering and the intermediate regime should be 
interpreted, we review the problem of electronic transitions induced by 
nuclear
motion since it is mathematically very similar.
In this problem, transition amplitudes are obtained from the conserved 
current. 
The same applies to quantum cosmology and indicates that 
probability amplitudes should based on the current
when  backscattering is neglected.
We then review why, in a relativistic context,
backscattering is interpreted as pair production whereas it is not in the non 
relativistic case. 
In each example
the correct interpretation is obtained by coupling the system to 
an external quantum device. From the absence of such external systems in 
cosmology, 
we conclude that  backscattering 
does not have a unique 
consistent 
interpretation in quantum cosmology.

\newpage 

\section{Introduction}\label{intro}

Quantum gravity suggests the possibility of non unitary evolution. In
particular singularities and regions of strong curvature, such as
black holes and the Big Bang, are natural places where such effects
could occur. However, in these cases,
the lack of comprehension 
of Planck scale 
physics precludes us from reaching any 
definitive conclusions.

More surprisingly,
when the universe is macroscopic and the curvature is small,
the possibility of non unitarity remains. 
Moreover, in this long wavelength regime, it seems unavoidable 
that quantum gravity implies the Wheeler-DeWitt (WDW) 
equation\cite{W}\cite{DW} 
(truncated to a finite set of modes 
so as to eliminate ultraviolet problems)
since this 
constraint expresses and guarantees reparametrisation
invariance. 

Thus the theoretical basis for investigating unitarity violation in quantum
cosmology seems well established. 
Nevertheless
 no definite analysis has been carried out, even in 
minisuperspace. Indeed, in
the literature one finds either prudent 
discourses which avoid the problem or claims ranging from
``there are no unitary
violations''\cite{GV} to ``there are finite violations of 
unitarity''\cite{vil}.

In order to understand the origin of the difficulties which have
hindered previous work on this question, it is necessary to 
recall how the usual unitary evolution of quantum matter in a given 
four-geometry is recovered from
the WDW equation. This has already been thoroughly
investigated\cite{GV}--\cite{MP}
and it is now well understood that {\it if} the wave
function is in a tight wave packet, i.e. the spread of matter energy is
small, and {\it if} gravity can be treated semiclassicaly, i.e. it can be 
described by a WKB wave function, then 
one should factor out this WKB
wave and interpret the remainder as the
wave function of matter which 
evolves according to the Schr\"odinger equation.

Therefore, when investigating the problem of unitary violations,
two problems must be confronted.
The first one is technical and 
concerns the precise mathematical characterization
of the corrections to the 
Schr\"odinger equation which are due to the 
abandonment of the two restrictions. 
In this respect, it should be mentioned that most
approaches have kept as starting point (i.e. as an ansatz) 
that the total wave function can be factorized into 
a gravitational wave characterizing 
the background and the rest which is still interpreted
as the matter wave function.
It is then very difficult to distinguish the
intrinsic corrections from the
those due to this factorization ansatz.

The second aspect is more fundamental and has to do with 
the interpretation of the wave function of the universe 
$\Psi(a,\phi)$, solution of the WDW equation  
($a$ designates the scale factor and 
$\phi$ matter variables in a minisuperspace context).
In particular, in order to address the question of possible violations of
unitarity, one must know how to extract the wave
function of matter when matter is not in a tight wave packet. 
Several proposals have been
made in the literature, see \cite{isham} for a review.
Two of the most specific ones are the conditional probability
interpretation wherein the
probability to find $\phi$ at $a$
is given by $ |\Psi(a,\phi)|^2 / {\int d\phi | \Psi(a,\phi)|^2}$
and the Vilenkin\cite{vil} interpretation which is based on the
current carried by $\Psi(a,\phi)$.

In this paper, by exploiting the 
generalized adiabatic treatment of \cite{MP},
we circumvent the difficulties which
have hindered previous work. The major advantage of this
treatment is to organize the WDW equation in such a way
that the corrections to the two approximations
can be separately and systematically analyzed before
adopting any interpretation scheme.
We then focus on the possibility of 
unitary violations in the light of
the mathematical properties of these corrections.
Our analysis will be restricted to minisuperspace models.
In a subsequent paper, we intend to extend it 
to cosmologies characterized by many variables. 

The first important conclusion of this work (which generalizes 
that of \cite{Par3}) is that when matter is not
in a tight wave packet but in the absence of backscattering
between waves with positive and
negative gravitational momentum (which correspond
to contracting and expanding universes),
the only consistent interpretation is that based
on the current. From our analysis, it will also be 
clear why when matter is in a tight wave
packet, i.e. when both approximations mentioned before 
are exact, different interpretations, such as the
probability or the current interpretations, become
equivalent. 

The second point of this article is to analyze the importance 
of the 
effects induced by backscattering
of gravitational waves.
We show that because of
destructive interference, the
amplitude for a forward wave 
to backscatter is 
proportional to 
$\exp(-{\partial_a p /p^2})$ where $p$ is the momentum of $a$. 
This is much smaller than previous estimates that suggested
that the amplitudes were 
proportional to the quantity 
$\partial_a p /p^2$
itself.

The third main conclusion concerns the consequences for matter
evolution of the couplings induced by backscattering. We show that 
even though
these transition amplitudes can be precisely evaluated,
 it is impossible to deduce what are 
exactly their physical consequences.
That is, quantum cosmology is an incomplete theory that
does not carry its own interpretation. Happily this incompleteness
only manifests itself at the level of backscattering and therefore
concerns exponentially small effects.

To reach these conclusions we 
address successively the mathematical and interpretational aspects. 
In the first part of the article (sections \ref{Potential} to 
\ref{NonUnit}), we carry
out an analysis of the WDW equation which does not make the
ansatz that matter should be in a tight wave packet. Dropping this
ansatz has already been advocated in 
\cite{kim1}\cite{wdwgf}\cite{wdwpt}\cite{Par3}. Here we shall 
follow the treatment of \cite{MP} (see also the earlier work of
\cite{DTK}) 
which does not require the
validity of the WKB approximation. Thus it can be used to obtain
quantitative estimates for both amplitudes to backscatter and 
effects of widely spread wave functions. From 
this treatment one unambiguously establishes that:
\begin{itemize}
\item
When 
backscattering amplitudes can be neglected and matter is
in a tight wave packet, matter obeys
the Schr\"odinger equation
in the background defined by the center of the wave packet,
thereby recovering the situation of refs. \cite{BV,vil,GV}.

\item
When matter is no longer in a tight wave packet, but 
backscattering amplitudes are
 still neglected, matter obeys a Schr\"odinger-like
equation wherein evolution is parameterized by $a$
and wherein gravitational back reaction effects 
are included at the quantum level, i.e. not only in the mean.
Unitarity follows from the ordered 
nature of the WKB propagation of $a$.

\item
When backscattering amplitudes are not neglected,
matter evolution is drastically modified
in the sense that matter states
associated with forward and backward 
propagating universes interact. This directly follows 
from the second order 
character
of the WDW equation in 
$\partial_a$.
\end{itemize}
These results follow from the separation of the 
length scales which govern the dynamics in quantum cosmology
when the universe is macroscopic.
The Compton wave length of $a$, $1/p$, is 
much smaller than both
the Compton wave length of a 
typical matter degree of freedom and than 
the time over which the geometry changes 
$(\partial_a p/p)^{-1}$.
%The smallest is $1/p$, the Compton wave length of $a$,
%the intermediate one is  the Compton wave length of a 
%typical matter degree of freedom
%state 
%and the largest is $\partial_a p/p$,
%the time over which the geometry changes.

In the second part of this article (sections \ref{INT} to \ref{beyond}) 
we address the problem of
interpretation. As basis for the discussion, we consider 
two physical examples whose mathematical description is 
similar to the WDW equation. These examples are electronic 
non adiabatic transitions
during atomic collisions and pair creation in an external potential.
The manner in which the wave functions should be interpreted in these
examples is standard text book material. 
Nevertheless, 
it is 
very instructive to delve in detail into the mathematical and
physical basis of these interpretations in order to see if they 
apply to quantum cosmology.

In both cases the conserved current serves as a basis for
extracting predictions from the theory. The basic mathematical reason
is that this is the only quantity which is conserved by virtue of the
equations of motion. In cosmology, in the absence of
backscattering, this procedure leads unambiguously to the
current interpretation.
In this derivation, the existence of well
 separated length scales plays also a crucial role:
It provides the possibility of asking
questions concerning light degrees of freedom
under the condition that the heavy one is 
propagating forward and located 
in a small neighborhood. 

The other important point is that 
when backscattering between the forward and backward propagating
waves occurs,
an {\it additional} principle  is required in order to  interpret  
backscattering amplitudes.
Moreover, this principle differs for the two examples.
In each case, it is dictated by the possibility of
coupling the system to a quantum external device which can act as
a measuring device.
In quantum cosmology,
no such external quantum device can be introduced, and it is therefore
impossible to decide 
which of these two interpretations, or possibly a
completely different third one, applies to the WDW equation. 
Hopefully the resolution of this problem 
can ultimately be obtained from a deeper
understanding
of the inner (ie. ultraviolet) structure of quantum gravity.

\section{Backscattering}\label{Potential}

In view of the importance of backscattering effects
in the problem of unitarity violation in quantum cosmology,
it is appropriate to understand clearly why backscattering
occurs and how to compute backscattering amplitudes.
To this end, we first consider the simple
problem of a universe in which 
matter stays in a given eigenstate $\ket{n}$
of constant energy $E_n$. 
This enables us to study the corrections to the
WKB approximation unencumbered by the details of matter
evolution. This analysis will serve as a guide to the next section
wherein 
the matter Hamiltonian no longer commutes with the WDW equation.
%matter evolution is nontrivial.

With this mathematical simplification, 
the WDW equation in minisuperspace reduces to one degree of
freedom moving in a potential 
\be
H_T \Psi = [G^2\partial_a^2 +V(a)+ 2a G E_n]\Psi =0
\label{One}\ee
where $G$ is Newton's constant.
The gravitational 
potential $V(a)$ comes from the curvature of space (for instance,
$V= -a^2 + \Lambda a^4$ in a closed universe with a cosmological
constant $\Lambda$).

The WKB solutions to eq. (\ref{One}) with unit Wronskian are
\be
\chi_n (a)= {e^{-i \int^a da' p_n(a')} \over \sqrt{2p_n} }
\label{wkb2}
\ee
and its complex conjugate, where 
$p_n(a) ={1 \over G}\sqrt{V(a)+ 2aGE_n}$ is the classical
solution of $H_T=0$ when matter has energy $E_n$. These
solutions correspond to expanding and contracting universes.

This identification 
relies on a sign convention which we now explain. First 
one expands (as usual when building wave packets) 
the phase in eq. (\ref{wkb2}) around a reference 
energy $\bar E$  to obtain
\be
\chi_n (a)\simeq {e^{-i \int^a da' \bar p_n(a')} \over \sqrt{2p_n} }
e^{-i \int ^a da' {\partial p \over \partial E} (E_n - \bar E)}
\quad .
\label{wkb2'}
\ee
Classical mechanics tells us that $ {\partial p 
\over \partial E}\vert_{E=\bar E} =
\pm {d \bar t \over da}$ where $\bar t(a)$ is the proper time 
obtained from the classical trajectory of $a$
driven by the reference matter energy $\bar E$.
The $\pm$ sign correspond to expanding and contracting universes
respectively. Thus we obtain
\be
\chi_n (a)\simeq {e^{-i \int^a da' \bar p_n(a')} \over \sqrt{2p_n} }
e^{\mp i \bar t(a) (E_n - \bar E)}
\quad .
\label{wkb2''}
\ee
If one adopts the usual convention that the phase dependence on
energy and time should be $e^{-i E_n t}$, then one must take the $-$ sign
in eq. (\ref{wkb2''}).
Hence $\chi_n$ corresponds to an 
expanding universe and $\chi_n^*$ to a contracting universe.
Obviously this choice is purely conventional and has
nothing intrinsic to it. The problem of defining an intrinsic arrow of
time in quantum cosmology is a delicate one, and we shall briefly
return to it at the end of this paper.

There are two kinds of corrections to the WKB solutions. The first are
corrections to the phase and norm of $\chi_n$ and $\chi^*_n$ separately. 
They are obtained by a standard calculation in which on writes
$\Psi = Ae^{iS/\hbar}$, inserts this ansatz
into eq. (\ref{One}) and calculates $A$ and $S$ as series in $\hbar$,
see for instance\cite{Migdal}.
The second kind of corrections
are due to the couplings between the forward and the 
backward solution.
They
describe the backscattering amplitudes engendered by the $a$ 
dependent effective potential $V_n(a)= V(a) + 2aG E_n$. 

We are interested in the second corrections since, as we shall later
see,
 these are the relevant ones for the problem of unitarity violations.
To analyze them we rewrite
$\Psi$ as
\ba
\Psi &=& {\cal C}_n (a) \chi_n(a) + {\cal D}_n (a) \chi_n^*(a)
\quad ,
\label{rel1}\\
\partial_a \Psi &=&  i p_n(a) \left[{\cal C}_n  (a) \chi_n (a) -  
{\cal D}_n (a)
\chi_n^*(a) \right] 
\quad .
\label{rel}\ea
The coefficients ${\cal C}_n $ 
and $ {\cal D}_n $ are uniquely determined by these
equations\cite{Baird} and
the conserved current (Wronskian) takes the simple
form
\be
\Psi^* i \lr {\partial_a} \Psi = 
|{\cal C}_n  (a)|^2 - | {\cal D}_n (a)|^2 = constant
\quad .
\ee
Differentiating eq. (\ref{rel1}) and comparing with eq. (\ref{rel})
yields 
\begin{equation}
\partial_a {\cal C}_n   \chi_n (a) +  \partial_a  {\cal D}_n 
\chi_n^*(a) -{\partial_a p_n \over 2 p_n}  
\left[{\cal C}_n  (a) \chi_n (a) + {\cal D}_n (a)
\chi_n^*(a) \right] =0
\quad .
\label{yeq}
\end{equation}
Inserting eq. (\ref{rel}) into  eq. (\ref{One}) 
and using eq. (\ref{yeq})  yields first order
coupled equations 
\ba
\partial_a {\cal C}_n  &=& 
{{\cal D}_n  \over 2} {\partial_a p_n  \over p_n } e^{-2i \int^a da\ p_n }
\quad , \nonumber\\
\partial_a  {\cal D}_n  &=& {{\cal C}_n  \over 2} {\partial_a p_n  \over p_n }
e^{+2i \int^a da\ p_n} 
\label{coupled}\ea
which
are equivalent to the original WDW equation  (\ref{One}).

$ $From eq. (\ref{coupled}) one reads that 
the instantaneous coupling 
between the forward and backward propagating waves is
proportional
to $ \partial_a \ln p_n $ times an oscillating phase. Thus, naively,
one would think that backscattering effects are of order $ 
\partial_a \ln p_n$. However when the effective potential 
$V_n(a)$
is slowly varying (i.e.
in the WKB limit $\partial_a  p_n/ p_n^2 \ll 1$)
this is incorrect because successively backscattered waves
destructively interfere. This destruction is so effective that
the backscattering amplitude is exponentially small.

To establish this, suppose that there are two asymptotic regions,
$a<a_-$ and $a > a_+$ wherein $V_n=const$ and  hence the WKB 
approximation is exact. Let us consider
the solution which  initially contains no backward
propagating wave: ${\cal D}_n(a < a_-) =0$ and ${\cal C}_n
(a <a_-) =1$. 
The backscattering amplitude is then ${\cal D}_n(a >a_+)$. It can be
calculated perturbatively by taking ${\cal C}_n=1$ in
eq. (\ref{coupled}) (for a more rigorous justification of this result see 
\cite{RBS,Migdal}):
\be 
B_n = {\cal D}_n
(\infty) \simeq \int 
da
\  {1 \over 2} {\partial_a p_n \over p_n} e^{+2i \int^{a} da'\ p_n(a')}
 \simeq e^{2i\int^{a^*}da' \  p_n(a')}
\quad .
\label{nomch}
\ee
The second approximation follows from the fact that
in the WKB limit, the integral 
is dominated by the (complex) saddle point $a^*$ where
$p(a^*)=0$. For instance when $V_n(a)$ has a
minimum at $a=a_0$: $V_n(a) = v_n + {V_n'' \over 2} (a-a_0)^2 +
O((a-a_0)^3)$, the saddle is located at $a^* \simeq a_0 \pm i \sqrt{2 v_n
\over V_n''}$ and the backscattering amplitude is
${\cal D}_n(+\infty) \simeq e^{- \pi v_n / 2G\sqrt{V_n''}}$
upon neglecting cubic and higher order terms in $a-a_0$.
The lessons from this exercise are that:
\begin{itemize}
\item
The naive estimate of the backscattering amplitude is incorrect. A
more detailed analysis shows that it is exponentially small.
\item
If the kinetic energy of gravity ($=V_n(a)/G^2$) 
has a minimum at $a_0$, then the
backscattering arises from a region of width $\Delta a = 
({1\over p_n} \partial_a^2 p_n)^{-1/2}$ around the minimum.

\end{itemize}

However, 
when the effective potential $V_n(a)$ 
does not have adequate asymptotic regions,
it is impossible to isolate unambiguously 
a backscattering amplitude.
Indeed, since the WKB approximation
is no longer asymptotically exact,
one needs an additional
principle to distinguish
 backward from forward classical motion. 
This situation generally arises
in cosmology. For instance, for a de Sitter universe, $V(a) = 
{-a^2 + \Lambda a^4} 
$ and backscattering around any  $a \gg 
\Lambda^{-1/2}$
far from the turning point 
results from interference over a distance $\Delta a 
\simeq a$. So it is impossible to isolate
the backscattering around $a$ from the effects due to the turning
point at $\Lambda^{-1/2}$. A similar situation obtains  for a matter
dominated universe $V_n(a) = -{a^2 } + {2 a G E_n}$: 
one cannot isolate the
backscattering around a given $a$ from the contribution of the origin
$a=0$ and the turning point $a= 2 GE_n$. In these cases, effects at the
turning points (for instance boundary conditions at $a=0$ or
$a=\infty$) might
 dominate any backscattering which occurs in intermediate regions.

\section{The Wheeler-DeWitt equation in minisuperspace}\label{Mini}

We now generalize the previous analysis to the case when the matter
Hamiltonian is not constant. Our aim is to find the equivalent of eqs. (\ref{coupled})
so as to be able to reveal the interplay between matter transitions
and backscattering effects.
We thus consider 
\begin{equation}
\left[G^2 \partial_a^2 +V(a) + 2aG \hat H_M(\pi,\phi,a) \right]\ket{\Psi(a)} =0
\label{WDW}
\end{equation}
where $\hat H_M$ is
the matter Hamiltonian operator. It depends on the matter coordinates $\phi$,
their conjugate momenta $\pi$ and $a$.
Notice that the wave function $\ket{\Psi(a)}$ is expressed as a 
ket only for the matter degrees of freedom.

Our analysis of the solutions of eq. (\ref{WDW}) is based on a double
hypothesis. First we require that the universe be macroscopic. 
This condition is satisfied if the radius of the
universe is larger then the Planck length
and if the total matter energy $\bk{H_M}$ is larger than the Planck mass.
This condition guarantees that the propagation of $a$
is WKB, i.e. ${d \ln p\over da}\ll p$. 
Secondly, we require that no individual matter 
quantum dominates the kinetic energy of gravity. 
That is we do not consider the case of an inflaton field
whose expectation value $\bk{\phi}$ is macroscopic.
Instead we are considering usual field configurations
such that $\bk{\phi} \simeq 0$ but with $\bk{H_M} \neq 0$
and macroscopic.

This second condition breaks the otherwise existing symmetry between
the radius of the universe $a$ and the matter degrees of freedom.
It ensures that $a$ is the heaviest and is therefore 
singled out to parameterize 
the {\it transitions} among neighbouring matter states.

%In practice the first condition is satisfied if the radius of the
%universe is much larger then the Planck length, and the second is
%satisfied if the mass and kinetic energy of each individual matter
%degree of freedom 
%is much smaller than the radius of the universe divided by the 
%Planck length squared $E =\sqrt{ M^2 + k^2} \ll { a\over G}$. Thus
%our universe could contain black holes so long as their Schwarzschild
%radius is much smaller then $a$.

We therefore postulate a neet separation of length scales: 
the Compton wavelength $1/p$ of the universe is much smaller
than both distance over which the momentum of gravity changes
$ (d \ln p / da)^{-1}$ (i.e. approximately the Hubble radius) and 
than $1/ (E_n - E_m)$, the time scale associated with
a typical matter transition.
We emphasize however that out treatment is exact,
i.e. we do not neglect corrections, we simply postulate that
they are small.

Since the matter transitions are the lightest, they should be
treated quantum mechanically. This is implemented by making an
adiabatic expansion for the matter states, 
following \cite{MP}\cite{DTK}.  On the other hand the radius of the
universe is the heaviest degree of freedom, and satisfies the WKB
condition, hence we shall  make a  WKB expansion for the
gravitational waves. This double expansion is the basic tool we use to 
analyze eq. (\ref{WDW}).

%In order to analyze the solutions of eq.  (\ref{WDW}), we shall make a
%double expansion following \cite{MP}\cite{DTK}. First  we 
%make an adiabatic expansion for
%the matter states. Secondly, as in eq. (\ref{rel1}), we make a  WKB expansion of the
%gravitational waves. This double 
%expansion is justified by the following hierarchy
%of three length scales.
%The largest is the distance over which the momentum of gravity changes 
%$ (d \ln p / da)^{-1}$ (i.e. approximately the Hubble radius). 
%The intermediate one is $1/E_n$, the Compton wave length of 
%a typical matter state.
%The smallest length is the Compton wavelength $1/p$ of gravity. 
%This hierarchy breaks the otherwise existing symmetry 
%between $a$ and $\phi$. $a$ is the heavier and therefore it is 
%singled out to parameterize 
%the evolution of lighter degrees of freedom.

The adiabatic
expansion for matter is realized by making the instantaneous
(i.e. at fixed $a$)
diagonalization of the matter Hamiltonian
\begin{eqnarray} 
\hat H_M(a) \ket{\psi_n(a)}&=& E_n(a) \ket{\psi_n(a)}
\quad ,
\nonumber\\
\scal{\psi_m(a)}{\psi_n(a)}&=& \delta_{m,n}
\quad .
\label{diag}
\ea
We combine this with the WKB expansion in the following way.
Let $p_n$ be the classical momentum if matter has energy $E_n(a)$,
\be
-G^2 p_n^2(a) + V(a) + 2aG E_n(a) =0
\quad .
\ee
Then we  generalize eqs. (\ref{rel1}, \ref{rel}) and decompose $\ket{\Psi}$ as
\ba
\ket{\Psi (a)}&= & \sum_n \left[ {\cal C}_n(a) {e^{-i \int^a p_n(a')
da' } 
\over
{\sqrt{2p_n(a)}}} + {\cal D}_n(a) {e^{+i \int^a p_n(a') da' } \over
{\sqrt{2p_n(a)}}}\right ] |\psi_n(a)\rangle \label{REL1}
\quad , \\
\partial_a \ket{\Psi(a)} &=& 
 \sum_n -i p_n(a)\left [{\cal C}_n (a) {e^{-i \int^a p_n(a') da' } \over
{\sqrt{2p_n(a)}}} - {\cal D}_n(a) {e^{+i \int^a p_n(a') da' } \over
{\sqrt{2p_n(a)}}}\right ] |\psi_n(a)\rangle
\quad .
\label{REL2}
\ea
These equations univocally define the ${\cal C}_n$ and ${\cal D}_n$
coefficients and
ensure that these coeficients are constant in the 
limit wherein both the WKB
approximation and the adiabatic approximation are
exact. Moreover, using these expressions, the conserved current
yields the simple expression:
\be
\bra{\Psi} i \lrpartial{a}\ket{ \Psi} = \sum_n  
|{\cal C}_n (a)|^2 - | {\cal D}_n  (a)|^2 = constant
\quad .
\label{wronskk}
\ee

Taking the derivative of eq. (\ref{REL1}) and comparing with
eq. (\ref{REL2}) yields a relation between ${\cal C}_n$, $ {\cal D}_n$
and their derivatives which generalizes eq. (\ref{yeq}).
Then inserting eq. (\ref{REL2}) into
the WDW equation (\ref{WDW}) and using this relation gives
\ba
\partial_a {\cal C}_n &=& \sum_{m \neq n}
\scal{\partial_a \psi_m}{\psi_n}
{p_n + p_m\over 2\sqrt{p_n p_m}}
e^{i \int ^a (p_n - p_m) da' }
 {\cal C}_m\nonumber\\
& &+{\partial_a p_n \over 2 p_n} 
e^{2i \int ^a p_n da } {\cal D}_n\nonumber\\
 & & + \sum_m \scal{\partial_a \psi_m}{\psi_n}
{p_n - p_m\over 2\sqrt{p_n p_m}}
e^{i \int ^a (p_n + p_m) da' }
{\cal D}_m
\label{central}\ea
and the same equation with $ {\cal C}_n \leftrightarrow {\cal D}_n$,
$i \leftrightarrow -i$. These equations are {\it{equivalent}} to the
original WDW equation. The essential advantage of this rewriting
is that it neetly separates the backscattering effects encoded in the 
last two terms from the matter transitions in the forward sector which 
are described by the first term.
As an important consequence, it enables the
WKB approximation to be implemented without factorizing,
as usualy done in former analysis, a gravitational wave
common to all matter states, i.e. without being obliged 
to make the hypothesis that 
matter is in a tight wavepacket in energy. 

\section{Unitary Evolution}\label{Unit}

When the WKB condition $\partial_a p/p^2 \ll 1$ is fulfilled,
there is no backscattering and one can correctly neglect the coupling
between the forward and backward propagating waves.
One thus obtains two uncoupled evolutions 
for forward and backward propagating universes. The equation
governing the forward sector is
\be
\partial_a  {\cal C}_n  = \sum_{m \neq n}
\scal{\partial_a \psi_m}{\psi_n}
{p_n + p_m\over 2\sqrt{p_n p_m}}
e^{i \int ^a (p_n - p_m) da }
  {\cal C}_m
\quad .
\label{centralU}
\ee
This equation 
has two nice properties\cite{MP}.
First it describes the
unitary evolution\footnote{
Some authors\cite{Peres} have advocated that the WDW equation should
be replaced by a first order equation in $\partial_a$ so as to
eliminate backscattering and guarantee unitarity.
In our setting, this would correspond to legislate that 
eq. (\ref{centralU}) is the ``correct'' equation governing
quantum cosmology.}
of matter as a function of $a$
and secondly it includes gravitational backreaction effects
at the quantum level, i.e. not only in the mean.

Unitarity directly 
follows from the fact that the right hand side of
eq. (\ref{centralU}) is antisymmetric. This can also be deduced
from eq. (\ref{wronskk})
and the fact that 
the coefficients ${\cal C}_n(a)$  and ${\cal D}_n(a)$  evolve
independently.
To see more explicitly how this
Schr\"odingerian character is obtained,
we define two 
effective matter wavefunctions associated respectively with the
expanding and contracting sectors
\ba
\ket{\phi_{ef\!f}^+(a)} 
&=& \sum_n { e^{-i \int ^a p_n da'} \over \sqrt{2 p_n} }
| \psi_n \rangle \langle \psi_n|  
\left[ e^{i \int ^a p_n da'} i\lrpartial{a} |\Psi (a)\rangle 
\right]
\quad ,\nonumber\\
&=& 
\sum_n  {\cal C}_n(a)  e^{-i \int ^a p_n da'} \vert \psi_n(a)\rangle
\nonumber\\
\ket{\phi_{ef\!f}^-(a)} &=& 
\sum_n  {\cal D}_n(a)  e^{+i \int ^a p_n da'} \vert \psi_n(a)\rangle
\quad .\label{phieff}
\ea
They obey Schr\"odinger type equations
\ba
\pm i \partial_a \ket{ \phi_{ef\!f}^\pm }= \hat H_{ef\!f}(a)
\ket{ \phi_{ef\!f}^\pm }
\label{Seff}
\ea
where the Hermitian operator
$\hat H_{ef\!f}$ plays the role of Hamiltonian. It is  defined by 
its instantaneous diagonalisation in the basis $|\psi_n(a)\rangle$:
\ba
\hat H_{ef\!f}(a)_{nm}
= \delta_{nm} p_n(a)
+ i \scal{\partial_a 
\psi_m}{\psi_n}
{(\sqrt{p_n} - \sqrt{p_m})^2 \over 2\sqrt{p_n p_m}}
\quad .\label{HEFF}
\ea
The peculiar form of the second term arises from the interplay between
the r.h.s. of eq. (\ref{centralU}) and the derivatives acting on  
the $|\psi_n(a)\rangle$
in the l.h.s of eq. (\ref{Seff})\footnote{Equation (\ref{HEFF}) corrects an
error in \cite{MP}}.

Thus, in the absence of backscattering,
 $\phi_{ef\!f}^\pm$ have  the following
properties: 
\begin{itemize}
\item $\phi_{ef\!f}^+$ is decoupled from $\phi_{ef\!f}^-$.
\item
$\phi_{ef\!f}^\pm$ are
local in the sense that  they depend only on $\ket{\Psi(a)}$ and
$\partial_a \ket{\Psi(a)}$ at $a$. 
\item
$\phi_{ef\!f}^\pm$ obey a linear equation, since they depend
linearly on the original ket $\ket{\Psi(a)}$. Thus they still 
obey the superposition principle. 
\item
Their norm is constant since
$H_{ef\!f}(a)$ is hermitian: 
\be\langle\phi_{ef\!f}^+(a)|\phi_{ef\!f}^+(a)\rangle 
= \sum_n |\langle
\psi_n(a)|\phi_{ef\!f}^+\rangle|^2 = \sum_n |{\cal C}_n(a)|^2 = const
\quad .
\ee
\item
If one works with wave packets tightly centered around 
a mean energy $\bar E$, 
 $\phi_{ef\!f}^\pm$ obey the time dependent Schr\"odinger equation
$i \partial_{\bar t} \phi_{ef\!f}^\pm = \hat H_M(\bar a_{\pm})  
\phi_{ef\!f}^\pm$
where
$H_M(\bar a_{\pm}(t))$ is the matter hamiltonian of eqs. (\ref{WDW},\ref{diag})
with $\bar a_{\pm}(t)$ describing respectively 
the classical expansion or contraction of the universe driven by $\bar E$.
To see this, as in eqs. (\ref{wkb2'}, \ref{wkb2''}), it suffices to develop the state 
dependent 
functions $p_n(a)$ around the mean value $\bar p(a)$.
To first order in $E_n - \bar E$, the second term of eq. (\ref{HEFF})
vanishes and the first one gives
\be
 \hat H_{ef\!f}(a)_{nm}\simeq
\delta_{nm} \{ (\bar p(a) + \partial_E \bar p(a) (E_n - \bar E) 
\}\quad .
\label{HEFF2}
\ee
Up to a term proportional to the identity 
(=$\delta_{nm} \{ \bar p(a) - \partial_E \bar p(a) \bar E\} $) which plays no
physical role, this diagonal matrix is identical to that defined 
by the matter Hamiltonian, see eq. (\ref{diag}),
since the overall factor $\partial_E \bar p(a) = \pm d\bar a_\pm/dt$
is precisely what is needed to convert $\partial_a$ in $\partial_{\bar t}$
in eq. (\ref{Seff}). 
 \item
 When one drops the tight wave packet approximation,
 $H_{ef\!f}$, defined in eq. (\ref{HEFF}), includes 
gravitational backreaction effects to the matter propagation
which depend on $n$. These are encoded in the non trivial dependence
of $p_n(a)$ on $E_n$
and in the second term on the right hand side.
Then, matter evolution must be parametrized by $a$
since it is meaningless to call upon a mean time parameter. 
\end{itemize}
In conclusion, we claim that $\phi_{ef\!f}^\pm$ defined in eq. (\ref{phieff})
are the unique functions of $\Psi$ which have these properties.
The only ambiguity lies in the definition of the WKB
approximation since one could modify 
$p_n$ in eq. (\ref{phieff}) by local terms of order $p_n ( 1 + O(d
\ln p_n /da))$ without modifying these properties. 
This ambiguity reflects the choice of normal ordering of the
operator $\partial_a^2$ in eq. (\ref{WDW}). 
However, these modifications are negligible
in the WKB regime.

Before discussing the role 
of backscattering effects, we wish to recall that
eq. (\ref{centralU}), the 
rewriting of 
is a convenient expression
when matter evolves quasi adiabatically,
that is when the distance scale over which matter makes transitions,
given by $\left( d \ln (E_n -  E_m) / da \right)^{-1}$,
is large compared to the wavelength $(E_n - E_m)^{-1}$. 
In this case one can easily calculate the non-adiabatic transition
 amplitude $A_{m \to n} $ 
from  $\ket{\psi_m}$ to $\ket{\psi_n}$. Moreover it 
proceeds along lines very similar to the computation of the
backscattering amplitude ${\cal D}_n$ in eq. (\ref{nomch}).
Indeed,
one first neglects transitions
and sets  ${\cal C}_n = \delta_{mn}$ for all values of $a$.
Then one inserts this zero-th order solution
 in eq. (\ref{centralU}) and integrates
over $a$ to obtain 
\ba
A_{m \to n} =
{\cal C}_n(a = + \infty) &\simeq&
\int da \scal{\partial_a \psi_m}{\psi_n}
{p_n + p_m\over 2\sqrt{p_n p_m}}
e^{i \int ^a (p_n - p_m) da }\nonumber\\
&\simeq & e^{i \int^{\tilde a} da' (p_n(a') - p_m(a'))}
\quad .\label{mtra}
\ea
In the second line, as in eq. (\ref{nomch}),
 we have evaluated the integral by saddle
point. The saddle is now at the complex value
$\tilde a$ solution of $p_n(\tilde a)= p_m(\tilde a)$. 

Since  $a (E_n - E_m)$
is much smaller than $G p_n^2$ (by virtue of the hierarchy
of length scales discussed  above),
one can expand
$p_n - p_m$ in powers of $E_n - E_m$. Keeping only the first term in
this series, one finds that the transition is given by the
usual expression controlled by the
imaginary time to reach the saddle point times the difference of
energy of the matter states, see \cite{MP} for further discussion of
matter transitions in quantum cosmology.

\section{Consequences of backscattering}\label{NonUnit}

To determine the range of validity of the truncated 
equation (\ref{centralU}), one must compute the importance 
of the effects induced by backscattering.
These effects arise when one abandons the 
WKB approximation. In this respect, it should be stressed
that there are two 
types of corrections to this approximation: local ones 
which can be evaluated by 
expanding the wavefunction as a series in $\hbar$ \cite{Migdal}, 
and global ones which mix forward and backward waves.
Only the latter lead to drastic modifications
of matter evolution. Thus, one should differentiate ``in the WKB
approximation'' from ``in the absence of backscattering''.
Our computation of backscattering amplitudes 
shall be based on
eq. (\ref{central})
and shall proceed as in eq. (\ref{mtra}) and eq. (\ref{nomch}).

Our first conclusion is that 
these amplitudes are exponentially small, whether or not matter
dynamics are neglected.
The new feature with respect to Section \ref{Potential} 
is that in eq. (\ref{central}) there are now coupling 
terms between ${\cal C}_n$ and all ${\cal D}_m$.
This leads to the fact that
the relative amplitude to
backscatter into different states is approximately thermal,
with the inverse temperature given by the twice the imaginary time needed to
reach the saddle point.
These backscattering transitions modify 
in a fundamental way
the structure of the evolution of the coeficients ${\cal C}_n$
since the evolution no longer closes onto itself. This implies two 
unusual features: first, an initially purely forward 
wave packet will inevitably leak out into the backward sector
and secondly, the knowledge of the initial values of the ${\cal D}_n$
is necessary to determine the evolution of the ${\cal C}_n$.

Let us first analyse the backscattering without change of matter state.
The second term of eq. (\ref{central})
has exactly the same form as the term studied 
in section \ref{Potential}. Thus it can be analysed in the same way: 
let us suppose 
that the effective potential $V_n(a)= V(a) + 2a G E_n(a)$ 
has a minimum at $a_n$ and that
we can approximate
$G^2 p_n^2 = V_n(a) \simeq V_n(a_n) + {V_n'' \over
2} (a - a_n)^2$.
Then, if the universe is purely forward propagating in state $n$ 
for $a<\!\!<a_n$, that
is ${\cal C}_m (a<\!\!<a_n) = \delta_{n m }$ and ${\cal D}_m  (a<\!\!<a_n) =
0$, the amplitude to backscatter into matter state $n$ is ${\cal
D}_n (a >\!\!> a_n)$. As in section \ref{Potential} this amplitude can be
perturbatively calculated by saddle point technique. 
The saddle is the solution of $p_n(a^*_n) =0$ and
it is located at $a^*_n \simeq a_n + i \sqrt{ 2 V_n(a_n) \over V_n''}$.
Upon neglecting higher order derivatives of $V_n(a)$, the
amplitude to backscatter is
\begin{equation}
B_{n \to n} = {\cal D}_n(a>\!\!>a_n)
\simeq \exp \left( {-{\pi V_n(a_n) \over G \sqrt{ 2 V_n''}}}\right) =
\exp \left(
{- { \pi \over 2}
p_n(a_n) {\rm Im}( a^*_n) }\right)\label{gamma}
\quad .\end{equation}
The quantity which appears exponentiated is 
the ratio of the distance over which $V_n(a)$
changes ($={\rm Im} (a^*_n)$) to the wavelength of the universe 
($=1/p_n$).
It is the ratio of the longest length scale in the universe to
the shortest. Therefore, 
$p_n(a_n) {\rm Im}( a^*_n) \gg (p_n(\tilde a)- p_m(\tilde a))
{\rm Im}( \tilde a)$, c.f.  
 eq. (\ref{mtra}), since the second expression is 
 controlled by the {\it difference} $p_n - p_m$.
Thus $B_{n \to n}$ is exponentially smaller then 
$ A_{n \to m}$.
It is this latter inequality which determines the validity of 
the first order equation (\ref{centralU}).

The backscattering with change of matter state is mediated by the
last term in eq. (\ref{central}). It can be evaluated in 
the same way and is given by 
\ba 
B_{n \to m} &\simeq& \int_{0}^{\infty}
da
\  \scal{\partial_a \psi_m}{\psi_n}
{p_n - p_m\over 2\sqrt{p_n p_m}}
 e^{+i \int^{a} da'\ (p_n(a')+p_m(a'))}
\nonumber\\
&\simeq& e^{i\int^{a^*}da' \  (p_n(a') +p_m(a'))}
\label{nomch2}
\ea
where $a^*$ is the saddle point solution of $p_n(a^*) +p_m(a^*)=0$.
To evaluate this saddle point integral, we define the mean energy
$\bar E(a) = { E_n + E_m \over 2}$ and expand the integrand to
first order in $E_n(a)- E_m(a)$:
$p_n(a') +p_m(a') = 2\bar p(a') + (E_n - E_m) {\partial_E \bar p }$.
This yields
\be
\vert B_{n \to m} \vert^2 
\simeq e^{- 4 {\rm Im} \left( \int^{a^*} da'\ (\bar p(a')\right) }
\times e^{ -2 \int_0^{{\rm Im}T^*}dt  
(E_n-E_m)} 
\ee
where $T^*= \int^{a^*}da'\ {\partial_E \bar p}$ is the 
time to reach the saddle point in the mean geometry.
Thus the relative distribution of the backscattered states is
approximately thermal with the extremely low temperature given by
$(2 {\rm Im}T^*)^{-1}$.

In the above we have neglected the fact that backscattering can
take place in several steps. For instance one can first backscatter
from matter state ${\cal C}_m$ to matter state ${\cal D}_{n'}$ and
then change matter state from ${\cal D}_{n'}$ to ${\cal D}_{n}$. Such
multi step transitions can compete with the direct
transition. Determining which channel is dominant is a difficult task
(see for instance \cite{HP}). Nevertheless a rough calculation of these
multi step backscatterings shows that the result based on the direct
channel gives a correct estimate: the amplitude to backscatter is an
exponentially small quantity governed by the total 
momentum $p_n$  and the backscattered states are approximately
thermally distributed.

Note that if the (unusual) sign of the kinetic term of gravity were
positive instead of negative, then the opposite would be true and
backscattering to states with higher matter energy would be
favored. Mathematically this change comes about because
$\partial_E p$ would have the opposite sign.

\section{Interpretation: introduction}\label{INT}

The literature on quantum cosmology abounds with 
interpretations of the wave function of the universe
$\Psi$. This reflects the contradictory views that are held on what
the wave function should describe and what it should predict.
However the question of interpretation is, to  a large
extent, determined by mathematical and physical consistency.

In order to develop a coherent interpretation we shall first
consider simple physical systems
--electronic non
adiabaticity in atomic or molecular collisions and particle creation in
external fields-- whose mathematical
formulation is extremely close to 
the WDW equation in minisuperspace models.
For these systems the interpretations, i.e. the procedures
 that must be used to
extract predictions from the theory, are known. 
They can thus serve as guides to suggest the interpretation
in quantum cosmology but also as
laboratories to formulate the restricted set of questions that 
we, living in the universe, can ask in quantum cosmology.
We then return to the WDW equation and compare  different
interpretations of $\Psi$ that have been proposed in the
literature.

The first conclusion of this investigation is that, 
upon neglecting backscattering,  
the current interpretation of Vilenkin appears to be singled out as
the only consistent one. Indeed, it leads to probability amplitudes 
which satisfy the usual properties of quantum amplitudes:
the superposition principle and decoherence of remote configurations
are both guaranteed. This is not the case if one adopts the conditional 
probability interpretation.  This interpretation instead 
does make sense for systems on which localized 
external devices can interact. 
Therefore it is inoperative in quantum cosmology since no such
external system can be introduced.
%on the basis that it does not
%respect the superposition principle. 
%the only 
%consistent interpretation is the current interpretation of Vilenkin. 
%Indeed, even though the conditional 
%probability interpretation does make sense for systems
%on which localized external devices can interact,
%it is ruled out in quantum cosmology on the basis that it does not
%respect the superposition principle. 

The second point concerns the physical interpretation 
of backscattering for the two examples. In each case, 
an additional principle, also based on the possibility of coupling 
the system to an external device,  is required 
to reach the correct interpretation. 
In cosmology, no such principle is available and therefore
the interpretation of backscattering remains ambiguous.
For instance whether or not one should ``third quantize'' 
and attribute some statistic to $\Psi$ 
would modify the physical consequences of the
backscattering amplitudes.

\section{Electronic non adiabaticity in atomic collisions}\label{elect}

Let us begin by considering the problem of two colliding atoms. After
factoring the center of mass coordinate and the total angular
momentum, the residual Hamiltonian looks like
\begin{equation}
H_{atom} = {P^2 \over 2 M} + V(R) + H_{el}(R, q_i,p_i) \label{HRel}
\end{equation}
where $R$ is the relative distance  between the two nuclei, $P$ the
momentum conjugate to $R$, and $q_i, p_i$ the electronic coordinates.
If we consider an energy eigenstate of this Hamiltonian 
\begin{equation}
\left[ - {\partial_R^2 \over 2M} +V(R) +  \hat H_{el}(q_i,p_i,R) \right]
\ket{\Psi_E(R)} 
= E\; \ket{\Psi_E(R)} 
\label{Sch}
\end{equation}
it has exactly the same structure as the WDW equation in
minisuperspace, see eq. (\ref{WDW}). 
Indeed in both cases there is a heavy degree of
freedom (the scale factor $a$ or the nuclear coordinate $R$) and
light degrees of freedom (the matter $\phi$ or the electrons
$q_i$). Because of this similarity the same techniques 
can be applied to both equations.

The simplest approximation consists in treating the coordinate 
$R$ classically as a given function of time: $R(t)$. Then the residual
degrees of freedom, the
electrons, propagate according to the time dependent equation
$i \partial_t \ket{\psi_{el}} = \hat H_{el} (R(t),q,p)  
\ket{\psi_{el}}$ and the 
non-adiabatic
dynamics are encoded in the transition amplitudes from one instantaneous
eigenstate to another one (Landau--Zener effect). 

If these transitions are too energetic or the motion of $R$ is
governed by a too spread out wave packet, 
the background field approximation is no longer correct.
Then one must solve eq. (\ref{Sch}) and 
the techniques discussed in Section \ref{Mini} should be
brought to bear\cite{DTK}. 
Upon neglecting backscattering 
effects but taking recoil effects into account, 
one obtains a second regime
in which the dynamics is governed by
the equivalent of eq. (\ref{Seff}): 
 Schr\"odinger like equations for the electrons in terms of the
amplitudes ${\cal C}_n(R)$ and ${\cal D}_n(R)$ associated respectively
with forward and backward WKB waves governing the inward
and outward motion of $R$. In these equations,
unitarity is respected for the ${\cal C}_n$ and ${\cal D}_n$
amplitudes separately and
the role of time is played by the heavy coordinate $R$. 
When backscattering effects in $R$ cannot be neglected, these effective 
Schr\"odinger equations are no longer valid and one obtains the third situation
wherein one must solve
the full equations for coupled ${\cal C}_n, {\cal D}_n$
system, the equivalent of eq. (\ref{central}).

The sole difference between the atomic problem and the 
WDW equation in minisuperspace 
is the sign of the kinetic energy of the
heavy nuclei which is positive whereas that of $a$
is negative. This difference only plays a role in the third situation
wherein it affects the identification of in and out modes
and the spectrum of backscattered states,
see the remark at the end of Section  \ref{NonUnit}.
As far as the light degrees of freedom are concerned,
the second regime in atomic physics and in cosmology
are completely analogous.

For this reason it is very instructive to 
review the interpretation of the solutions of
eq. (\ref{Sch}). Indeed, the aspects which are often glossed over
in text books are precisely the ones needed 
for the interpretation of quantum cosmology. 
In particular, we shall discuss two compatible but nevertheless
different schemes of interpretation. In the first, 
one restricts the analysis
to asymptotic amplitudes to find the system in
a given state. In the second scheme,
it is the wave function itself which is 
interpreted at all times. Upon investigating the 
relationship between these schemes, the
separation of the length scales governing electrons
and nuclei will again play a crucial role.

The first interpretation, proposed by Born\cite{Born},
concerns scattering amplitudes.
In the present case as well,
eq. (\ref{Sch}) describes a
scattering event and the main question one wants to answer (and to
compare to experimental data) is how
the amplitude of finding a certain final state depends on the initial
state of the atom.
Let us therefore suppose that the atoms are initially in their ground
state and approaching each other, i.e. ${\cal C}_n (R=+ \infty) = \delta_{n0}$.
According to Born, the probability
for the atoms to end up in state $n$ is $|{\cal D}_n(R= + \infty)|^2$.
The mathematical basis for this identification 
is the conservation of the Wronskian  $  \int dq_i 
\Psi_E^* i \lrpartial{R} \Psi_E $. Indeed since the nuclei 
repel
each other the region $R<0$ is inaccessible and one must impose
$\Psi_E(R=0)=0$. Hence the Wronskian vanishes everywhere (since it
vanishes at $R=0$). 
Therefore if one expands $\Psi_E$ in terms of 
forward and backward coefficients
${\cal C}_n$ and ${\cal D}_n$ as in section \ref{Mini}, the vanishing
of the Wronskian implies that
$\sum_n |{\cal C}_n(R)|^2 = \sum_n |{\cal D}_n(R)|^2$. 
In particular if $\sum_n |{\cal C}_n(R=+\infty)|^2 =1$, then $
 \sum_n |{\cal D}_n(R= + \infty)|^2 = 1$ which expresses the conservation of
probability during the scattering process.

In order to deal with dynamically induced backscattering
(which occurs in quantum cosmology)
it is necessary to consider situations in which the wave
function is not restricted to vanish at a certain place.
For instance one can think of a single atom
moving in one dimension in a static but inhomogenous
electric field. The Schr\"odinger
equation for the atom has the same form as eq. (\ref{Sch}), but now
$R$ is the position of the atom and 
the $R$ dependence of $V(R) + H_{el}(R,q,p)$ 
is due to the electric field. Consider an atom
incoming from $R= - \infty$ in a given energy eigenstate $n_0$. 
This is implemented 
by requiring that the solution satisfy the following boundary
conditions $|{\cal C}_{n_0}(R = -
\infty)|^2 =\delta_{n, n_0}$.
Since there is no atom incoming from $R= + \infty$,
one should also impose ${\cal D}_n (R= + \infty) =0$ for all $n$.
By these boundary conditions one specifies what are
the in-modes, i.e. modes which possess a well defined 
semi-classical behavior in the remote past. 
In this determination, there is an identification
of the asymptotic waves carrying a unit Wronskian
with the corresponding classical
trajectories obtained by the stationary phase condition
applied to wave packets or by identifying directly
the wave length times $\hbar$ with the momentum
of the particle.

Upon solving the full (untruncated) Schr\"odinger equation one obtains
both the asymptotic
amplitudes for the atom to continue over
the potential barrier, ${\cal C}_n (R= + \infty)$,
as well as those to be reflected by the potential
${\cal D}_n (R= - \infty)$. Unitarity is expressed by the equality of
the ingoing current to the sum of the outgoing currents:
\be
 \int dq_i 
\Psi_E^* i \lrpartial{R} \Psi_E =
1= |{\cal C}_{n_0}(R = -
\infty)|^2 = \sum_n |{\cal D}_n (R= - \infty)|^2 + \sum_n |{\cal C}_n
(R= + \infty)|^2 
\quad .
\label{unittt}
\ee

We emphasize that in Born's interpretation
the identification of the quantities 
${\cal D}_n ,{\cal C}_n (R= \pm \infty)$ as scattering amplitudes, and
hence unitarity,
follow from the internal structure of
eq. (\ref{Sch}) and the use of asymptotic
waves normalized to unit Wronskian for all $n$. An external concept
was only necessary to identify the asymptotic incoming and outgoing
modes
(ie. the left and right movers).

The essential simplification of this interpretation is that one 
considers only asymptotic states. This leads naturally to the 
question whether one can also give an interpretation to
the electronic dynamics in the intermediate region. The coefficients
${\cal C}_n(R)$ and ${\cal D}_n(R)$ are natural candidates for this
interpretation since they generalize to finite $R$ the coefficients
that where the basis for the asymptotic interpretation. 
Moreover, this is also 
supported by the fact that, when backscattering can be neglected, they
obey a Schr\"odinger like equation with a Hamiltonian that tends in
the limit of infinitely heavy nuclei to the time dependent electronic
Hamiltonian $H_{el}(R(t),q,p)$.

To further investigate this question, we now turn to the second
interpretation of the solution of eq. (\ref{Sch}). In this
interpretation --which for reason of simplicity is generally the first one
presented in text books--, $|\Psi_E(R,q)|^2$ is interpreted as the
probability for the atom to be at $R$ in electronic configuration $q$.
The basis for this interpretation is two fold. First if one considers
time dependent solutions $\Psi(t,R,q)$ of the Schr\"odinger equation,
then there is a conserved current
\begin{equation}
\partial_t |\Psi(t,R,q)|^2 + \partial_R J_R  + \partial_q J_q =0
\end{equation}
where
\begin{equation}
J_R = \Psi^* i \lrpartial{R} \Psi \quad , \quad 
J_q = \Psi^* i \lrpartial{q} \Psi
\quad .\end{equation}
Hence it is consistent to interpret $|\Psi(t, R,q)|^2$ and therefore
$|\Psi_E(R,q)|^2$ as probability
densities, see {\it e.g.} \cite{Mes}.
The second basis, due to von Neumann\cite{vonNewmann}, 
is the following:  if we couple the
atom to an idealized measuring device (initially in the state $X_R=0$,
$X_q=0$) through a coupling
$\delta(t) | q \rangle| R \rangle P_{X_q} P_{X_R} 
\langle q |\langle R | $ where
$P_{X_R} $ and $P_{X_q}$ are the  
variables conjugate to $X_R$ and $X_q$, the measuring device will
record outcomes $R,q$ (i.e. $X_R=R$ and $X_q=q$) 
with probability $|\Psi(t, R,q)|^2$.

An essential point to note about these two arguments is that they both
necessitate the introduction of external concepts which are
used in situ and no longer only asymptotically: in the first case,
external time, and in the second, the local measuring device. For this reason
the application of this second interpretation 
will be problematic in quantum cosmology since
these external concepts cannot be invoked.

It is now very instructive to show that
this interpretation --when used with care--
is consistent with the S matrix
interpretation. In particular, we shall show when and why
 the second interpretation also implies  that the coefficients
${\cal C}_n (R)$ and ${\cal D}_n(R)$ should be interpreted as the
amplitudes for the electrons to be in state $n$ when the nucleus 
are propagating forward and backward at (better near) $R$. In this, the 
necessity of having well separated  length scales is crucial.

Consider the probability $|\Psi_E(R,q)|^2$ that the atom is at $R$ and
the electrons in state $q$. In terms of the coefficients 
${\cal C}_n (R)$ and ${\cal D}_n(R)$ it takes the form
\begin{equation}
|\Psi_E(R,q)|^2 =  \sum_n \vert \left( {\cal C}_n (R) { e^{i \int^R p_n dR} \over
\sqrt{ 2 p_n(R)}} + {\cal D}_n(R)  { e^{-i \int^R p_n dR} \over
\sqrt{ 2 p_n(R)}} \right) \langle q|\psi_n(R)\rangle \vert^2
\quad .
\label{normm}
\end{equation}
This quantity is wildly oscillating because of the interference terms
between ${\cal C}_n (R)$ and ${\cal D}_n(R)$. These
interferences arises from the
localized measuring device which 
has interacted strongly with both the electrons
and the nuclei. Since we are interested only in the electrons, we
would like the measuring device to interact weakly with the
nuclei. 
Thus we consider a measuring device that interacts with the nucleus
over a certain range of $R$, and with a certain momentum
sensitivity. That is we consider the probability that the nucleus
be in a wave packet
$\varphi_{R_0,P_0} = { e^{-(R-R_0)^2/ \Delta^2} e^{i R P_0}
}/\sqrt{2 \pi \Delta}$ where as an example we have taken
a Gaussian packet. If the interval $\Delta$ satisfies both
$\Delta \gg 1/p_n$ and $\Delta \ll  (\partial_R \ln {\cal C}_n )^{-1}$,
the probability amplitude to find the nuclei in state 
$\varphi_{R_0,P_0}$ is approximately
\begin{equation}
\int\! dR \varphi^*_{R_0,P_0}(R)  \Psi_E(R,q)
\simeq
\sum_n  {\cal C}_n(R_0) \langle q | \psi_n(R_0)\rangle
{e^{-(p_n(R_0) -P_0)^2 \Delta^2} \over \sqrt{ 2 p_n}}
\label{interC}
\end{equation}
where we have neglected the $R$ dependence of
${\cal C}_n(R) \langle q | \psi_n(R)\rangle$ over the interval
$\Delta$, and dropped the exponentially small contribution of the
${\cal D}_n$ coefficients.
Thus only the amplitudes with momentum in the interval $P_0 \pm 1/
\Delta$ are selected. From 
this expression it follows that 
\be
{| {\cal C}_n(R_0) |^2 \over 2 p_n(R_0)}
\label{Csqrt}
\ee
 is the probability that the
electrons are in state $n$ if the nuclei 
are in the interval $R_0 \pm \Delta$.
Recalling we are working at fixed total energy $E$ 
and that $p_n(R_0)$ is proportional to 
the speed of the nuclei,
it follows that we have calculated is the probability for the
electrons to be in state $n$
multiplied by the time they spend in the
interval $R_0 \pm \Delta$. This confirms  the interpretation of
${\cal C}_n(R)$ as the probability\footnote{
For the same reason in
relativistic theory one must remove $1/\sqrt{2 \omega}$ from
scattering matrix elements in order to get probability amplitudes.
This is known as the reduction formula.}
 amplitude for the electrons to be in state $n$.

In this derivation the
identification of ${\cal C}_n(R)$ as electronic amplitudes 
is inevitably approximate. Indeed eq. (\ref{interC}) 
followed from the fact that the wavelength of the nucleus and that of the
electrons are very different.
The same separation of length scales was also
the justification for neglecting the coupling between forward and
backward propagating modes in eq. (\ref{central}), thereby obtaining a
Schr\"odinger equation for the forward propagating coefficients
only. Thus, this separation is twice used 
in order to reach the interpretation of ${\cal
C}_n(R)$ as the amplitude for the light
degrees of freedom to be in state $n$.

We now turn to the consequences of backscattering.
In this case,
$\sum_n |{\cal C}_n(R)|^2$ is no longer constant. How is this interpreted
in the context of electronic non adiabaticity?

To address this question, consider
a generic solution of eq. (\ref{Sch}) in the context of the atom
propagating in a background field so that the whole real axis $-\infty < R <
+\infty$ is accessible. For 
such a solution, 
none of the coefficients
${\cal C}_n(\pm \infty)$, ${\cal D}_n(\pm \infty)$ are zero.
Using 
the equality
\be
\sum_n |{\cal C}_n(+\infty)|^2 +
\sum_n |{\cal D}_n(-\infty)|^2 =
\sum_n |{\cal C}_n(-\infty)|^2 +
\sum_n |{\cal D}_n(+\infty)|^2
\ee
the probability to be in state $n$ and forward propagating at
$R=+\infty$ is
\be
P({\cal C}_n(+\infty)) = { |{\cal C}_n(+ \infty)|^2 \over 
\sum_n |{\cal C}_n(+\infty)|^2 +
\sum_n |{\cal D}_n(-\infty)|^2 }
\ee
when the WKB approximation is asymptotically exact.
This probability is a highly nonlocal concept since it involves the
coefficients
${\cal C}_n$ and ${\cal D}_n$ at $+$ and $-$ infinity
respectively. For this reason it is a relevant and useful
concept only for an external observer who has
 access to both the forward and backward waves 
at $R= \pm \infty$. However, there are many relevant 
quantities which are local in $R$.
Moreover these quantities 
govern the physical outcomes when one is in the 
second regime wherein backscattering is neglected.
In addition light subsystems can presumably only ask questions
concerning these quantities.
An example of such a local quantity is
the relative probability to be forward propagating in state $n$ or state
$m$ at $R= +\infty$:
\be
{ P({\cal C}_n(+\infty)) \over P({\cal C}_m(+\infty))}
=  { |{\cal C}_n|^2 \over |{\cal C}_m|^2 }
\label{Csqrt2} \quad .
\ee

In such local quantities the absolute normalization
involving a mixture of ${\cal C}_n$ and ${\cal D}_n$
coefficients disappears. 
However they do not obey a closed equation since the
${\cal C}_n$ are coupled to the ${\cal D}_n$.
Thus,
the effect of the backscattering is to modify the effective Schr\"odinger
equation for the forward propagating coefficients ${\cal C}_n(R)$
into a stochastic equation, where the stochasticity is given
by the absence of knowledge a forward propagating subsystem has about
the backward propagating part of the solution. At the end of the
calculation he must average over the unknown coefficients ${\cal D}_n$.
Happily, when the electronic length scales are well separated
from the nuclear length scales, this stochasticity is exponentially small.

\section{Relativistic particle in an external
field}\label{relativistic}

In order to further investigate the relationships
between the conservation of the Wronskian, the identification
of the asymptotic solutions and the interpretation of the
wave functions, another situation should be considered: 
the Klein-Gordon (KG) equation governing the
propagation of a relativistic particle in an external field. 
This example has in common with the WDW equation 
a Lorentzian signature,
i.e. the quadratic form determining the kinetic energy term is
not positive definite. 

This analogy has been emphasized by many authors,
see e.g. \cite{MTW} for a discussion at the classical level,
and has been used to advocate the necessity of performing 
a ``third quantization''. 
Indeed, in the case of the relativistic particle,
unitarity is implemented by carrying out a ``second quantization''. 
The difference with the non relativistic case
is that the Wronskian is no longer
interpreted as a particle current density but as a charge density. This
follows from a different  identification
of the in and out asymptotic modes which in turn
dictates the new interpretation.

Let us review these features by considering
the propagation of a charged particle in an external electric field. 
We shall take for simplicity the electric field to be homogeneous and
to point in the $x$ direction. We impose that the 
electric field vanishes for $t\to
-\infty$ and for $t\to +\infty$ which provides us with asymptotic
regions in which the momentum is constant and therefore the modes are 
exactly WKB. 
The Klein Gordon equation in 
the gauge $A_t =0$, $A_x = f(t)$, $A_y=A_z =0$ takes the form
\begin{equation}
\left[\partial_t^2 - \left( \partial_x + ieA_x(t)\right) - \partial_y^2 -
\partial_z^2  + m^2 \right]\Psi =0
\quad .
\label{KG}
\end{equation}
Writing  $\Psi =e^{i (k_x x + k_y y + k_z z)} \phi_k(t)$, one finds that
$\phi_k$ obeys the equation
\begin{equation}
\left[\partial_t^2 + V_k(t)\right]\phi_k =0
\label{KG2}\end{equation}
where $V_k(t) = (k_x + e A_x(t))^2 + k_y^2  + k_z^2   + m^2$.

This equation is identical to that studied in section \ref{Potential}
and can be analyzed in the same way.
To reach an interpretation one first needs to identify the in-mode 
describing a particle incoming from $t= - \infty$. 
This is achieved by decomposing $\phi_k $ as
${\cal C}_k(t) \chi_k(t) + {\cal D}_k(t) \chi_k^*(t)$,
where $\chi_k(t)$ is the WKB solution with unit positive Wronskian
of eq. (\ref{KG2}).
The in modes are then given by imposing the boundary
condition $|{\cal C}_k(t=-\infty)|^2 = 1, {\cal D}_k(t=-\infty)
= 0$. 
(Notice that the condition on the coefficient ${\cal D}_k$ is applied on the 
same ``side'' of the potential in opposition to what is
done in the non-relativistic case).
As in the non-relativistic case, the potential $V_k(t)$ 
induces  backscattering effects and
both the coefficients ${\cal C}_k$ and ${\cal D}_k$ are non
vanishing at $t=+
\infty$.
The conservation of the Wronskian 
now reads  
\be
1 =  \int d^3 x
\Psi^* i \lrpartial{t} \Psi =
|{\cal C}_k(t=-\infty)|^2 
= |{\cal C}_k(t= +\infty)|^2 - |{\cal D}_k(t
= +\infty)|^2 
\quad .
\ee
How should one interpret these coefficients at $t=+\infty$, and in
particular the fact that ${\cal C}_k(t=+\infty) >1$? This is the content
of the Klein paradox that was discussed in the early days of
relativistic quantum field theory\cite{Klein}.

Probably the simplest way to reach a physical understanding
is to proceed as in the non-relativistic case by building wave packets 
(in $k$) and following the classical trajectories they describe.
One then finds that the wave packets made of ${\cal C}_k \chi_k$
follow, both for early and late times, the trajectories of a positive
 charged particle. But the part of wave packets proportional to
 ${\cal D}_k \chi_k^*$ follow trajectories of negatively charged
 particles. Thus the conservation of the Wronskian expresses 
charge
 conservation and the decomposition into ${\cal C}_k$ and ${\cal D}_k$ is a
decomposition into positive and negative charged particles. 
Hence the extra charge carried by the ${\cal C}_k(t= +\infty)$
is compensated by the presence of 
anti-particles that reach $t= +\infty$:  $|{\cal D}_k(t
= +\infty)|^2 > 0$. 
Another more mathematical way to reach this conclusion is to note that
eq. (\ref{KG}) is invariant under gauge transformations, and that the
Wronskian is the charge associated to the gauge field $A_\mu$.

Once this is realized one still faces the problem of extracting
(probabilistic) predictions from the theory. This is done by second
quantizing $\Psi$: the coefficient ${\cal C}_k$ becomes an operator
destroying a particle of momentum $k$.
In this framework, one establishes that in addition 
to the ``elastic'' scattering of initial (anti) particles, one is 
describing pair creation. Even in the in-vacuum state, pairs of particles 
are spontaneously created with probability  amplitudes
governed by the backscattering amplitude
 $B_k = {\cal D}_k(t= +\infty)$ calculated in eq. (\ref{nomch}), which in
this context are called Bogoljubov coefficients.
If some particle is initially present, 
one finds that there is also induced emission.
In this reasoning the conclusion that the Wronskian is expressing
charge conservation relied on an external input, namely that
$t$ in eq. (\ref{KG2}) 
is an external time which allowed to follow asymptotic motion.
Moreover, the prediction that many pairs can be produced
can be checked experimentally.

We now discuss how these new effects modify the dynamics of {\it
internal} degrees of freedom.
The most natural way to implement internal degrees of freedom 
is to go back to eq. (\ref{KG}) and take the momenta along the
$x,y,z$ directions to be the additional degrees of freedom.
The model becomes non trivial when the electric field is not
homogeneous, so that $k_x, k_y, k_z$ are not constants of motion (they
can be replaced by adiabatic constants of motion as in section
\ref{Mini}).
The choice of how the field should
be second quantized is dictated by the spin-statistics
theorem\cite{spinstat}, ie. $\Psi$  is either a bosonic or a fermionic
field.
The important conclusion for us is
that these different second quantization procedures are not equivalent,
even if one restricts oneself to the dynamics of the internal degrees
of freedom.

To second quantize one first
selects a complete set of modes with only positive frequency at
$t=-\infty$,
$\phi_k^{in}$ to which one associates a particle destruction operator
$a_k^{in}$. To the complex conjugate of these modes one associates an
antiparticle destruction operator $b_k^{in}$. If we neglect
backscattering then the number operators $N_{part} = \sum_k a_k^{in
\dagger} a_k^{in}$ and $N_{anti-part}= \sum_k b_k^{in
\dagger} b_k^{in}$ commute with the second quantized Hamiltonian.
The Fock space then decomposes into non interacting sectors. The
simplest sector 
is the vacuum $|0\rangle$. The first non trivial states are
the one particle (or antiparticle) states $\sum_k f(k) a^{in
\dagger}_k |0\rangle$. These states are exact analog's of the forward
propagating modes in the atomic example of the previous section, and
should be interpreted in the same way.
The next states are the two particle states
$\sum_{k,k'} f(k,k')  a^{in
\dagger}_k  a^{in\dagger}_{k'} |0\rangle$ where $f$ is symmetric or
antisymmetric according to the statistics of the field. 
Since $f$ does not
factorize into a product of a function of $k$ and $k'$, the
interpretation of this state cannot reduce to that 
of one particle states. Moreover the statistics play a role in 
the determination of the final amplitudes since 
the presence of a particular quantum modifies the spectrum of
transition amplitudes through either Bose enhancement
 or Pauli suppression. Whether or not this has 
an interpretation from the point of view of an
internal degree of freedom is unclear. 

The situation is worse 
when one takes backscattering into account. In this case, the number
operators no longer commute with the Hamiltonian 
because of the non-vanishing
character of the Bogoljubov  coefficients $B_{k\to k'}$ (calculated in
eq. (\ref{gamma}) and (\ref{nomch2})) relating in and out modes.
This implies that in order to determine the amplitude to find a given
state $k$ in the one particle sector, it is necessary to know all the
initial amplitudes (in Fock space) to find particles and
antiparticles. Thus the amount of initial data needed in order to
extract exact predictions from the theory in the one particle sector
is (infinitely) larger than in the non relativistic case. For this
reason third quantization is not a very attractive alternative since
it diminishes the predictivity of quantum cosmology. Furthermore the
additional ingredients which imposed second quantization, namely the
presence of an absolute time and of a gauge field coupled to $\Psi$
are not present in quantum cosmology. For these reasons it seems
advisable, in the absence of additional strong arguments in favor of
it, not to resort to third quantization in quantum cosmology.

\section{The conditional probability interpretation
based on the norm of $\Psi$}

In the light of the previous examples we now return to the
interpretation of the WDW equation.

For many authors, conservation of probability is a
necessary requisite for quantum cosmology to possess a coherent
interpretation. 
The most widely adopted scheme 
which implements 
from the start unitarity and conservation of probability 
is the conditional
probability interpretation\cite{PW}\cite{GV}\footnote{
Although \cite{GV} does not use the name ``conditional
probability interpretation'', it implicitly adopts it. Indeed 
in \cite{GV} the decomposition of the wave function into a gravitational
and a matter wave is uniquely fixed
by imposing 
that the matter wave satisfies
$\int d\phi
\chi_s^*(a,\phi) \chi_s(a,\phi) = const$. Hence $\chi_s(a,\phi)$ is
identical to $\tilde \phi_{ef\!f}(a,\phi)$ defined in eq. (\ref{cp}).

Another recent paper that claims that matter evolves unitarily in
quantum cosmology \cite{KIM} contains mathematical errors. Indeed
$\psi(h_a)$ in eq. (7) must carry an index $n$ since the equation it
obeys depends on $n$ (see \cite{kim1} for a discussion of this $n$
dependence). But this implies that the second term on the right hand
side of eq. (10) is non hermitian. Furthermore the third term
proportional to $\Omega$ is non hermitian (it would only be hermitian
if integrated over $h_a$). Hence the conclusion
of \cite{KIM} that $\sum_n |c_n|^2$
is constant is incorrect.}
In this interpretation one identifies 
\begin{equation}
\tilde \phi_{ef\!f}(a,\phi) = 
{\Psi (a, \phi)  \over \sqrt{ \int d \phi \Psi (a,
\phi)^* \Psi(a, \phi)}}
\label{cp}\end{equation}
as the amplitude of probability to find matter in state $\phi$ at radius
$a$. By construction the norm square of $\tilde \phi_{ef\!f}$ is constant,
independently of the validity of the WKB approximation
or any other condition.

The main problem with this construction is that $\tilde \phi_{ef\!f}$,
contrarily to $ \phi_{ef\!f}^\pm$ defined in eq. (\ref{phieff}),
does not obey a linear equation (since $\Psi$ does). Therefore
it does not satisfy the superposition principle. 
This can be ignored as long as one considers 
tight wave packets such that the 
factorization of a single gravitational wave,
common to all matter states, offers a good approximation.
However, once the spread in matter energy is large,
the probabilities significantly 
vary even in the total absence of interactions.
To see this 
suppose that 
the matter Hamiltonian has only two 
instantaneous eigen-states $\ket{\psi_0}$ and $\ket{\psi_1}$. 
Then, in the WKB regime, the wave function of the universe is 
\be 
\ket{\Psi(a)}
= {\cal C}_0 {e^{-i \int^a da p_0(a)} \over \sqrt{2 p_0(a)}}\ket{\psi_0}
+  {\cal C}_1 {e^{-i \int^a da p_1(a)} \over \sqrt{2 p_1(a)}}\ket{\psi_1}
\quad .
\ee
Let us further suppose that it is impossible to make transitions from
state $0$ to state $1$. Then the 
 instantaneous states are truly stationary: $\partial_a \ket{\psi_{0 }
}=\partial_a \ket{\psi_{1 }     }
 =0$ and 
the coefficients ${\cal C}_0$ and $ {\cal C}_1$ are constant.
Using the conditional probability interpretation, one then finds that the
probability for finding matter in state $\ket{\psi_0}$  is
\begin{equation}
P(\psi_0 | a)= 
{ |{\cal C}_0|^2
 \over  |{\cal C}_0|^2 + {p_0 (a) \over p_1(a)}
|{\cal C}_1|^2} 
\label{CCC}
\quad .
\end{equation}
It reduces to the usual expression $P_0 (a)= |{\cal C}_0|^2$
only  when $p_0 (a) /p_1(a) = 1$, i.e. when $E_0 = E_1$. 
Therefore $P_0 (a)$ varies once $E_0 \neq E_1$
through the ratio of the two gravitational momenta 
$p_0$ and $p_1$ entangled with the two matter states
by the constraint.
The more remote these energy are, the bigger 
the effect since the 
associated momenta will be more different.
Moreover, upon taking into account backscattering effects, 
the probability receives in addition to the continuous component a 
rapidly oscillating component with frequency $2 p_0$
whose origin are the interference between expanding and contracting
solutions.

In our opinion, these two properties are sufficient to reject 
the conditional probability interpretation.
Before explaining our reasons, it is interesting to recall why 
a similar construction can be meaningful for a non-relativistic atom.
In that case, the probability to find the atom in
electronic state $\phi$ under the condition that it be at $R$ is
indeed  given by eq. (\ref{cp}). What does give a meaning to this expression
is the possibility (and the necessity)
of coupling the atom to a localized external
device so heavy that its own recoils be negligible.
Only then does the factor $p_0/p_1$ in eq. (\ref{CCC}) take
meaning as the relative time each state interacted with the detector,
and only then
can the forward and backward waves
give rise to the interferences at frequency $2 p_0$. 
But if one couples the atom to a measuring device
 that does not
significantly disturb the motion of the nuclei (see section \ref{elect}), 
one obtains the current interpretation based on eq. (\ref{Csqrt2}) 
instead that based on the norm\footnote{
Some authors have proposed to introduce an additional 
variable $\eta$ such that the modified 
WDW equation reads $i\partial_\eta \Psi = H \Psi$,
thereby recovering a situation similar to that of
an atom in a time dependent state. 
It is then very tempting to interpret $|\Psi(a, \phi, \eta)|^2$
as the probability to find $a$ and $\phi$ at $\eta$. However,
since in cosmology one must restrict oneself 
to questions such that only little momentum 
is transferd to $a$, one recovers the current
interpretation of eq. (\ref{Csqrt2}) (see section \ref{elect}). Thus this 
latter interpretation naturally arises irrespectively of 
the use of an additional  variable as time.}, see eq. (\ref{normm}).

To present our arguments against eq. (\ref{CCC})
we first recall the following point.
In quantum mechanics, from the solutions of the 
linear equation determining evolution 
(which can be either the Schr\"odinger or the WDW equation)
one obtains that 
very remote states neither interact nor interfere. From 
this mathematical property
and the conventional interpretation of probability amplitudes,
one then gets that the probabilities of such remote states
do not influence each other.
But this fundamental decorelation of remote 
configurations does not obtain in eq. (\ref{CCC}). 

From an 
epistemological point of view, the fact that linearity of 
the WDW equation does not reflect itself in a simple property of the
physical probabilities $|\tilde \phi_{eff}|^2$ is disturbing. Indeed if
linearity does not manifest itself in a simple way at the level of
observation, why insist that we start with a linear equation?  We might 
as well introduce non linearity at the level of the WDW equation directly.

%Therefore, it goes against the basic idea of
%decoherence in quantum mechanics
%according to which the evolution of non interfering states decorelate
%completely.
In our opinion, the only way to save this interpretation probably requires
to first restrict the applicability of eq. (\ref{cp}) to neighboring states only. 
But then its status would be of an approximate 
character rather than a starting point from which
all solutions of the WDW equations can be interpreted.

\section{Vilenkin interpretation within 
the WKB regime}

Of the different interpretations of the wave function of the
universe which have been proposed, the closest to the
physical
examples presented above is that of Vilenkin\cite{vil}, as generalized 
in \cite{Par3}. 
Indeed the essential lesson of these examples is that they give
physical substance to the coefficients ${\cal C}_n$ and ${\cal
D}_n$.
This is not surprising since these coefficients 
have the following mathematical properties:
\begin{itemize}
\item 
The conserved current is expressed most simply 
in terms of these coefficients upon working with 
modes normalized to unit Wronskian,

\item
they obey a linear first order equation in $a$,

\item
if backscattering is neglected they evolve independently 
and unitarily. 
\end{itemize}
In the absence of backscattering,
these properties designate the 
${\cal C}_n$ and ${\cal D}_n$ as the probability amplitudes
for matter to be in state $n$ in the forward or backward sector. 
We invite the reader to compare this reasoning with
the original one of Max Born, see \cite{Born}.

Very important is also the fact that in this interpretation, 
decoherence effects arise naturally and obey usual properties,
in contradistinction to what occurs in the conditional 
probability interpretation. For instance,
when two states or two groups of states have only vanishing 
matrix elements of $H_{ef\!f}(a)$, their 
evolution is completely independent. 
It is thus fully legitimate to interpret the 
wave function composed of their sum as representing 
two different universes rather that a quantum superposition of states
living in the same universe. This decoherence is the most basic one
(when compared to that induced by the environment or measuring
devices) since
it is solely determined by the nature of the full Hamiltonian,
i.e. the WDW constraint.
It therefore provides a natural explanation for the fact that 
one should not consider quantum superpositions
of states giving rise to different semi-classical geometries (histories)
since these states completely decouple.

The neglection of quantum backscattering effects
can also be conceived as a manifestation of decoherence.
Indeed, the appropriate character of the decomposition
into forward and backward solutions (rather than 
in sines or cosines) follows from the effective
decoupling of these two sets. Thus, there is
a hierarchy of decoherence associated with the 
hierarchy of length scales. Indeed, the  
decoupling of the forward and backward sectors
is the most efficient since backscattering amplitudes 
are exponentially smaller than matter transition amplitudes.
This leads to a clear separation of the solutions 
into two sets characterized by the sign of $p_n(a)$.

Thus the only aspect which differs from that 
presented in the physical examples of Sections 
\ref{elect} and \ref{relativistic} 
is the identification of these sets with forward and 
backward propagation in time. In the previous examples,
it was necessary to use an external time to identify
which asymptotic modes correspond to forward motion
at $t=- \infty$. (Recall that this identification was different 
for the non-relativistic and relativistic examples.)
In cosmology, an external time does not exist and thus 
the eventual determination of forward  motion in time,
i.e. the arrow of time, should be intrinsic.
Presumably, the answer is 
thermodynamic: if the matter is out of equilibrium, then the arrow of
time is determined by the direction in which entropy is increasing.
This point of view was recently advocated in the quantum cosmological context
in \cite{HLL}. If this is indeed the case, quantum cosmology is 
incomplete since it can only be interpreted for macroscopical 
universes containing enough degrees of freedom so that 
thermodynamics applies. For such universes, the hierarchy of 
matter and gravity scales that we twice exploit
will be well established. Therefore it appears that only
macroscopic universes can be meaningfully investigated
and interpreted.

\section{The interpretation in the presence of
backscattering}\label{beyond}

As emphasized by Vilenkin, since the WKB approximation is not
exact, {\em ``probability'' and
``unitarity'' are inherently approximate concepts}. Indeed 
neither the norm of $\phi^+_{ef\!f}$ nor of $\phi^-_{ef\!f}$ are 
separately conserved in the presence of
backscattering.

However this is not an insurmountable problem
for the following reason. Any subsystem
of the universe, for instance the degrees of freedom within 
the cosmological horizon, is necessarily an open system and its dynamics are
therefore necessarily non unitary. The coupling between forward and
backward propagating universes will appear to this subsystem as an
additional coupling to the environment. Furthermore this coupling
gives rise to exponentially small effects, with appearing in the
exponential approximately 
the ratio of the Hubble radius to the wavelength of the
universe. Hence it is completely negligible compared to the other
couplings to the environment at least for macroscopic universes.

Having said this, one may nevertheless want to know what the possible
effects of backscattering could be --however small they are--, 
and how they should be interpreted.

A first point to be noticed is that in most realistic cosmologies,
backscattering cannot be dissociated from effects at the origin of
the universe. The reason being that the distance
$({d\ln p \over d a})^{-1} $ over which the backscattering occurs is
comparable to the distance to the origin of the universe. 
For this reason backscattering and its interpretation may be
intimately linked to the ultra violet structure of quantum gravity.
If this is the case, 
the interpretation cannot be discussed in the context 
of a truncated WDW equation.

Nevertheless one may imagine universes in which asymptotic regions
exist. Then backscattering events can be localized and their
amplitudes properly evaluated. In this case, one can inquire
into their interpretation even in the context of simplified WDW
equations.
As discussed in sections \ref{elect} and \ref{relativistic} 
there  are at least two
consistent interpretations of backscattering.
In both cases the choice of interpretation was dictated 
by reinserting 
the system into a wider context. In the cosmological context 
we do not know
into what (if any) wider context the WDW equation fits, and thus 
the precise way backscattering should be interpreted is unclear. 

Let us nevertheless briefly discuss  the consequences of backscattering.
For a subsystem of the universe which is forward propagating, the
effect of backscattering is included by averaging over the (unknown)
state of the rest of universe. In a first quantized context, the rest
of the universe would simply be the backward propagating waves. In the
second quantized context it would correspond
to all the other sectors of the theory
with different ``universe and anti-universe numbers''. Thus the
averaging, and hence the effect of the averaging, is different in both
cases. 

Moreover, the interpretation of backscattered waves
themselves is problematic.
For instance what is the physical principle which 
would determine the arrow of time in the part of the 
wave function whose origin is backscattering ?
To be concrete, consider a solution of the WDW equation which for
$a \ll a_0$ contains only ${\cal C}_n$ coefficients, but for $a \gg a_0$
contains both ${\cal C}_n$ and ${\cal D}_n$ coefficients.
 ($a_0$ is the center of the region where the 
backscattering occurs). From section \ref{NonUnit} we know that
 the ${\cal D}_n(a)$ coefficients will ultimately be more or less thermally 
distributed with a very low temperature.
In which direction does the arrow of time point for the ${\cal D}_n$
coefficients? Is $a \simeq a_0$ the origin or the end of the universe?

In conclusion the separation between forward and backward propagating
universes, and hence the appearance  of an effective Schr\"odinger
equation for matter appears to be a more robust concept than
previously thought. Indeed the corrections to the Schr\"odinger
equation are exponentially smaller than any matter transitions. 
On the other hand the precise way
the backscattering manifests itself does not seem to be constrained by
the WDW equation --at least in the simple models we have considered in
this paper. Hopefully this interpretational problem can be resolved
once we have  
deeper understanding of the structure of quantum gravity. The
inclusion of anisotropies and inhomogeneities, and more importantly of
the ultraviolet sector of the theory, could lead to a unique consistent
interpretation.

{\bf Acknowledgments.}

S.M. is a ``chercheur qualifi\'e du FNRS''. He would like to
thank the Institute of Theoretical Physics at 
Utrecht University where part of this work was carried out.


\begin{thebibliography}{999}

\bibitem{W} J. A. Wheeler, in {\it 
Battelles Rencontres: 1967 Lectures on Mathematical Physics}, edited
by C. De
Witt and J. Wheeeler,  Benjamin,
New York, 1968 

\bibitem{DW} B. DeWitt, Phys. Rev. {\bf 160} (1967) 1113

\bibitem{GV} C. Bertoni, F. Finelli, G. Venturi,
Class. Quant. Grav. {\bf 13} (1996) 2375.

\bibitem{vil} A. Vilenkin, Phys. Rev. D39 (1989) 1116


\bibitem{banks} T. Banks, Nucl. Phys. B249 (1985) 332

\bibitem{BV} R. Brout and G. Venturi, Phys. Rev. D39 (1989) 2436


\bibitem{hartle} J. B. Hartle, in ``Gravitation and Cosmology'' edited by B.
Carter and J. Hartle, Plenum Press (1986)

\bibitem{isham} C. J. Isham, {\it Canonical Quantum Gravity and the Problem
of the time}, In ``Integrable Systems, Quantum 
Groups and Quantum Field Theories''
Kluwer Academic Publishers, London 1993, gr-qc/9210011

\bibitem{kiefer2} C. Kiefer, {\it The semiclassical approximation
to quantum gravity}, in ``Canonical Gravity-from Classical to Quantum''
ed. J. Ehlers and H. Friedrich (Springer, Berlin 1994), gr-qc/9312015

\bibitem{kim1} S. P. Kim, Phys. Rev. D {\bf 52} (1995) 3382 


\bibitem{wdwgf}R. Parentani, Nucl. Phys. B {\bf 492} (1997) 475

\bibitem{wdwpt}R. Parentani, Nucl. Phys. B {\bf 492} (1997) 501

\bibitem{Par3} R. Parentani, 
Phys. Rev. D {\bf 56} (1997) 4618 


\bibitem{MP} S. Massar and R. Parentani, 
Nucl. Phys. B {\bf 513} (1998) 375


\bibitem{DTK} J. B. Delos, W. R. Thorson and S. K. Knudson,
Phys. Rev. {\bf A 6} (1972) 709

\bibitem{Migdal} A. D. Migdal, {\it Qualitative Methods in Quantum
Theory}, Addison Wesley, 1989

\bibitem{Baird} L. C. Baird, J. Math. Phys. {\bf 11} (1970) 2235


\bibitem{RBS} V. L. Pokrovskii, S. K. Savvinykh, and F. R. Ulinich, 
Sov. Phys. JETP {\bf 7} (1958) 879; V. L. Pokrovskii and
I. M. Khalatnikov, Sov. Phys. JETP {\bf 13} (1961) 1207 

\bibitem{Peres} A. Peres, {\it Critique of the Wheeler-DeWitt
equation}, gr-qc/9704061

\bibitem{HP} J. Hwang and P. Pechukas, J. Chem. Phys. {\bf 67} (1977) 4640


\bibitem{Born} M. Born in {\it Quantum Theory of Measurement} ed. by
J. A. Wheeler and W. Zureck, Princeton University Press, Princeton,
New Jersey (1983)


\bibitem{Mes} Messiah, A. {\it Quantum Mechanics}, North-Holland (1969)

\bibitem{vonNewmann} J. von Neumann, {\it Mathematical Foundations of
Quantum Mechanics}, Princeton Univ. Press. Princeton 1955


%\bibitem{PageHawking} S. W. Hawking, D. N. Page, Nucl. Phys. B {\bf
%298} (1988) 789 

\bibitem{MTW} C. W. Misner, K. S. Thorne and J. A. Wheeler, 
{\it Gravitation}, Freeman, San Fransisco, (1973)





\bibitem{Klein} C. Itzykson and J.-B.
   Zuber, {\it Quantum Field Theory},  McGraw-Hill (1980)

\bibitem{spinstat} R. F. Streater and A. S. Wightman, 
{\it PCT, Spin and Statistics, and all that},
Benjamin (1964)

\bibitem{PW} D. N. Page and W. K. Wootters, Phys. Rev. D {\bf 27}
(1983) 2885

\bibitem{KIM} S. P. Kim, Phys. Rev. D 55 (1997) 7511. 


\bibitem{HLL} S.W. Hawking, R. Laflamme, G.W. Lyons, Phys. Rev. D {\bf
47} (1993) 5342


%\bibitem{HH} J. Halliwell and S. W. Hawking, Phys. Rev. D {\bf 31}
%(1985) 1777



%\bibitem{MG} W. H. Miller and T. F. George, J. Chem. Phys. {\bf 56} (1972)
%5637

%\bibitem{LMO} G. Lifschytz, S. D. Mathur, M. Ortiz, 
%Phys. Rev. D {\bf 53} (1996) 766 

%\bibitem{BerryBalazs} M. Berry and * Balazs, J. Phys. {\bf A 12}
%(1979) 625

%\bibitem{Zurek} 
%S. Habib, K. Shizume,
%W. H. Zurek, Phys. Rev. Lett. {\bf 80} (1998) 4361


%\bibitem{DDP} A. M. Dykhne, Sov. Phys. JETP {\bf 14} (1962) 941;
%J. P. Davis and P. Pechukas, J. Chem. Phys. {\bf 64} (1976) 3129


%\bibitem{Bfa} R. Parentani, 
%{\it The validity of the Background Field Approximation},
%gr-qc/9710059, to appear in the proceedings of the conference ``The Internal 
%Structure of Black Holes and Space-time Singularities'' 
%held in Technion, Haifa
%(1997)

%\bibitem{LZ} L. Landau, Phys. Z. Sowjetunion {\bf 2} (1932) 46;
%C. Zener, Proc. R. Soc. (London) {\bf A 137} (1932) 696;

%\bibitem{S} E. C. G. Stuckelberg, Helv. Phys. Acta {\bf 5} (1932) 369

%\bibitem{Berry} M. V. Berry, Proc. R. Soc. {\bf A 392} (1984) 45

%\bibitem{Berry2} M. V.  Berry, Proc. R. Soc. {\bf A 414} (1987) 31

%\bibitem{Jackiw} R. Jackiw, Int. J. Mod. Phys. {\bf A 3} (1988) 285

%\bibitem{SS} D. S. Salopek and J. M. Stewart, 
%Class. Quantum Grav. {\bf 9} (1992) 1943




\end{thebibliography}
\end{document}